# Geometric and disorder – type magnetic frustration in ferrimagnetic "114" Ferrites: Role of diamagnetic $Li^+$ and $Zn^{2+}$ cation substitution.

Tapati Sarkar, V. Caignaert, V. Pralong and B. Raveau*

*Laboratoire CRISMAT, UMR 6508 CNRS ENSICAEN,
6 bd Maréchal Juin, 14050 CAEN, France*

Dedicated to Professor Jacques Friedel on the Occasion of His 90$^{th}$ Birthday.

**Abstract**

The comparative study of the substitution of zinc and lithium for iron in the "114" ferrites, $YBaFe_4O_7$ and $CaBaFe_4O_7$, shows that these diamagnetic cations play a major role in tuning the competition between ferrimagnetism and magnetic frustration in these oxides. The substitution of Li or Zn for Fe in the cubic phase $YBaFe_4O_7$ leads to a structural transition to a hexagonal phase $YBaFe_{4-x}M_xO_7$, for M = Li ($0.30 \leq x \leq 0.75$) and for M = Zn ($0.40 \leq x \leq 1.50$). It is seen that for low doping values i.e. x = 0.30 (for Li) and x = 0.40 (for Zn), these diamagnetic cations induce a strong ferrimagnetic component in the samples, in contrast to the spin glass behaviour of the cubic phase. In all the hexagonal phases, $YBaFe_{4-x}M_xO_7$ and $CaBaFe_{4-x}M_xO_7$ with M = Li and Zn, it is seen that in the low doping regime (x ~ 0.3 to 0.5), the competition between ferrimagnetism and 2 D magnetic frustration is dominated by the average valency of iron. In contrast, in the high doping regime (x ~ 1.5), the emergence of a spin glass is controlled by the high degree of cationic disorder, irrespective of the iron valency.

* Corresponding author: Prof. B. Raveau
e-mail: bernard.raveau@ensicaen.fr
Fax: +33 2 31 95 16 00
Tel:  +33 2 31 45 26 32

## 1. Introduction

Strongly correlated electron systems, involving transition metal oxides with a "square" crystal lattice, namely perovskites, have been the subject of numerous investigations during the last thirty years, for their superconducting properties in cuprates as well as their magnetic properties in CMR manganites. Oxides with a triangular lattice have also been studied, as shown for the spinel family, [1 – 3] that exhibits strong ferrimagnetism and unique magnetic transitions, as for example in $Fe_3O_4$, and for the pyrochlore family and for spinels with a pyrochlore sublattice, [4 – 6] which have been investigated for magnetic frustration. Nevertheless, the number of oxides with a triangular lattice is much more limited, and the recent synthesis of the "114" cobaltites and ferrites [7 – 11] with original structures closely related to the spinel offers a new playground for the investigation of the competition between magnetic ordering and geometric frustration in this class of materials [12 – 14]. The consideration of the structure of the "114" ferrites, cubic $LnBaFe_4O_7$ (**Fig. 1 a**), and hexagonal $CaBaFe_4O_7$ (**Fig. 1 b**), shows that these oxides, both consist of similar layers of $FeO_4$ tetrahedra, called triangular (T) and kagomé (K), and that their "$Fe_4O_7$" frameworks can be deduced from each other by a translation of one triangular layer out of two. Therefore, the corresponding iron sublattice is very different in the two structural families: it consists of a pure tetrahedral framework of corner sharing "$Fe_4$" tetrahedra in the cubic $LnBaFe_4O_7$ (**Fig. 2 a**), whereas it is built up of rows of corner – sharing "$Fe_5$" bipyramids running along $\vec{c}$, interconnected through "$Fe_3$" triangles in the (001) plane for hexagonal $CaBaFe_4O_7$ (**Fig. 2 b**). It is this geometry of the iron framework which is at the origin of two different kinds of geometric frustration. The pure tetrahedral iron sublattice of $LnBaFe_4O_7$ oxides is similar to that of pyrochlore [4] and consequently generates a 3 D magnetic frustration. In contrast, the mixed "bipyramidal – triangular" sublattice of $CaBaFe_4O_7$, allows a competition between a 1 D magnetic ordering along $\vec{c}$ and a 2 D magnetic frustration in the (001) plane.

Recently, we have shown the possibility of stabilizing the iron "bipyramidal – triangular" lattice at the cost of the tetrahedral lattice by substitution of zinc for iron in the cubic $LnBaFe_4O_7$ oxides [15]. Surprisingly, it was observed that this doping with a diamagnetic cation destroys the spin glass behaviour of $LnBaFe_4O_7$ and induces ferrimagnetism. However, with progressive increase of the Zn concentration, the ferrimagnetic interaction starts to weaken, and we get a spin glass for very high doping concentration. In order to understand the role of the different factors which govern the magnetic properties of these ferrites, we have studied the substitution of lithium and zinc, two diamagnetic cations with different valencies, for iron, in the ferrites YBaFe$_4$O$_7$ and CaBaFe$_4$O$_7$. We discuss, herein, the relative influence of valence effects and cationic disordering upon the competition between magnetic ordering and frustration in these systems. We will specifically attempt to decouple the role of the Fe valency, and that of the disorder on the Fe sites in order to determine how the two separately affect the magnetic ground state. This will allow us the possibility to tune and customize the magnetic properties of these oxides by understanding the role of the two governing factors – the average Fe valency, and the degree of disorder on the Fe sites.

## 2. Experimental

All the samples used in this study were prepared by standard solid state reaction technique. The details of the synthesis procedure can be found in our earlier publications [15, 16]. The samples were chemically monophasic, and the phase purity was checked from X-ray diffraction patterns registered with a Panalytical X'Pert Pro diffractometer. The d. c. magnetization measurements were performed using a superconducting quantum interference device (SQUID) magnetometer with variable temperature cryostat (Quantum Design, San Diego, USA). The a.c. susceptibility, $\chi_{ac}$(T) was measured with a PPMS from Quantum Design with the frequency ranging from 10 Hz to 10 kHz ($H_{dc}$ = 0 Oe and $H_{ac}$ = 10 Oe). All the magnetic properties were registered on dense ceramic bars of dimensions ~ 4 × 2 × 2 mm$^3$.

## 3. Results and discussion

**3.1.** *Zn substitution in YBaFe$_4$O$_7$ and CaBaFe$_4$O$_7$*

In our previous work [15], we had synthesized the oxide series YBaFe$_{4-x}$Zn$_x$O$_7$ with the hexagonal symmetry, for x ranging between 0.4 – 1.5. For the sake of relevant comparison, we have prepared CaBaFe$_{4-x}$Zn$_x$O$_7$ samples with x = 0.5 and 1.5. We have specifically chosen these two values of x so that we can investigate how the two factors, i.e. the average Fe valency and the cationic disorder, affect the magnetic properties in two separate regimes: the low doping regime and the high doping regime. The XRPD patterns of these two samples clearly show that they are monophasic, keeping the hexagonal symmetry of CaBaFe$_4$O$_7$, and with cell parameters close to those of the virgin oxide i.e. *a* = 6.3527 (1) Å and *c* = 10.3274 (2) Å for x = 0.5, and *a* = 6.3668 (1) Å and *c* = 10.2975 (1) Å for x = 1.5.

### 3.1.1. *Low doping regime: $YBaFe_{3.5}Zn_{0.5}O_7$ and $CaBaFe_{3.5}Zn_{0.5}O_7$*

$YBaFe_{3.5}Zn_{0.5}O_7$ and $CaBaFe_{3.5}Zn_{0.5}O_7$ fall in the low doping regime. The degree of disorder (measured in terms of the % of substituent cation) is the same and relatively small in these two samples. We show the d. c. M vs T for the two samples in **Fig. 3**. Both samples show a ferrimagnetic transition (seen as a sharp rise in the M vs T curves), similar to that observed for $CaBaFe_4O_7$, [9] in accordance to the fact that the samples have been stabilized in the hexagonal symmetry. However, their ordering temperatures, and importantly, their magnetic moments are much smaller than those observed for $CaBaFe_4O_7$ ($T_C$ = 270 K and $M_{FC(5K)}$ = 2.6 $\mu_B$/f.u.). Moreover, the magnetic moments and the ordering temperatures for the two samples are very different, though they exhibit the same degree of disorder. The $CaBaFe_{3.5}Zn_{0.5}O_7$ sample exhibits a much higher magnetic moment ($M_{FC(5K)}$ = 1.7 $\mu_B$/f.u.) and transition temperature ($T_C$ ~ 203.0 K) compared to the $YBaFe_{3.5}Zn_{0.5}O_7$ sample for which $M_{FC(5K)}$ = 0.54 $\mu_B$/f.u. and $T_C$ ~ 119.5 K. Bearing in mind that both these oxides exhibit the same "bipyramidal – triangular" iron sublattice (**Fig. 2 b**), these results show that the competition between the 1 D ferrimagnetism that appears along $\vec{c}$ and the 2 D frustration in the (001) plane of the hexagonal structure is strongly affected by the average valence of iron. Indeed, for the same degree of disorder (12.5 % Zn), $T_C$ decreases and the magnetic frustration increases significantly as the average value of Fe decreases from 2.57 for $CaBaFe_{3.5}Zn_{0.5}O_7$ to 2.29 for $YBaFe_{3.5}Zn_{0.5}O_7$.

The effect of the higher value of the average Fe valency of $CaBaFe_{3.5}Zn_{0.5}O_7$ is also seen in the M vs H loops of the two samples (shown in the inset of **Fig. 3**). Not only is the coercivity and remanence magnetization higher for $CaBaFe_{3.5}Zn_{0.5}O_7$, the shape of the M-H loops are also very different. $CaBaFe_{3.5}Zn_{0.5}O_7$ has a square loop, reminiscent of hard ferrimagnets, while $YBaFe_{3.5}Zn_{0.5}O_7$ has a much softer M-H loop signifying a weakening of the ferrimagnetic interaction in $YBaFe_{3.5}Zn_{0.5}O_7$ compared to that in $CaBaFe_{3.5}Zn_{0.5}O_7$. *Thus, in the low doping regime, the average Fe valency of the ferrite is clearly the governing factor that controls the magnetic state and the strength of the magnetic interaction. The degree of disorder here plays a relatively minor role.*

**3.1.2.** *High doping regime: $YBaFe_{2.5}Zn_{1.5}O_7$ and $CaBaFe_{2.5}Zn_{1.5}O_7$*

An increase of the doping concentration of the diamagnetic substituent subsequently leads to the appearance of spin glass behaviour in both $YBaFe_{4-x}Zn_xO_7$ as well as $CaBaFe_{4-x}Zn_xO_7$. This can be seen in **Fig. 4**, where we have shown the d. c. M vs T curves for $YBaFe_{2.5}Zn_{1.5}O_7$ and $CaBaFe_{2.5}Zn_{1.5}O_7$. In contrast to the sharp rise in the M(T) curves below the ordering temperatures seen in the low doped samples (x = 0.5), the M(T) curves of the higher doped samples (x = 1.5) show a more gradual rise in the magnetization with the decrease in temperature terminating in a cusp-like behaviour at low temperature. The temperature at which the ZFC M(T) curves of the two samples show cusps are $T_{cusp}$ = 35.5 K and 40.5 K for $YBaFe_{2.5}Zn_{1.5}O_7$ and $CaBaFe_{2.5}Zn_{1.5}O_7$ respectively. A. C. susceptibility measurements $\chi'(T)$ of the two samples measured using different frequencies in the range 10 Hz – 10 kHz (**Fig. 5**) show that both samples show similar frequency dependent peaks with $T_g$ = 45 K and 50 K for $YBaFe_{2.5}Zn_{1.5}O_7$ and $CaBaFe_{2.5}Zn_{1.5}O_7$ respectively. The two samples also have very similar narrow S – shaped loops with almost the same values of the coercivity and remanence magnetization (see inset of **Fig. 4**). Moreover, the important point to note here is that the average Fe valency of these two compounds is very different ($Fe_{val}$ = 2.40 and 2.80 for $YBaFe_{2.5}Zn_{1.5}O_7$ and $CaBaFe_{2.5}Zn_{1.5}O_7$ respectively). In spite of such a large difference in the Fe valency, the two compounds behave strikingly similar to each other. We explain this result as the dominating role of the cationic disorder in samples where the degree of disorder is large. *Thus, for higher doped samples, the degree of disorder plays the deciding factor in stabilizing the magnetic ground state, and the Fe valency plays only a minor role.*

Remarkably, the magnetic behaviour of these highly doped hexagonal phases is very similar to that of the virgin cubic sample $YBaFe_4O_7$, which was shown to be a spin glass, with a rather similar $T_g \approx 50$ K [10]. Thus, a high degree of cation disordering in the hexagonal phase has an effect similar to the pure geometric frustration of the tetrahedral iron sublattice (**Fig. 2 a**) of the cubic phase i.e. it allows a complete magnetic frustration to be reached.

**3.2.** *Li substitution in $YBaFe_4O_7$ and $CaBaFe_4O_7$*

In the previous sections, we have shown that the magnetic properties of hexagonal Zn substituted "114" ferrites $YBaFe_4O_7$ and $CaBaFe_4O_7$ can be tuned on the basis of two doping regions – the low doping regime and the high doping regime. While the average Fe valency plays a crucial factor in determining the magnetic state of the oxide in the low doping regime, the degree of cationic disorder becomes the dominant factor controlling the magnetic state in the regime of high doping. In order to confirm that this effect is not specific to Zn, but is, in fact, a rather general phenomenon, we carry out similar studies on $YBaFe_4O_7$ with a second diamagnetic substituent, $Li^+$, and we compare the magnetic behaviour of $YBaFe_{4-x}Li_xO_7$ with that previously observed for $CaBaFe_{4-x}Li_xO_7$ [16].

Due to its univalent character, lithium has the advantage of inducing an average iron valency which is different from that of the zinc compounds for the same substitution rate, thereby allowing the relative effects of the Fe valence and the cationic disorder to be compared further. The size of $Li^+$, which is similar to that of $Zn^{2+}$, and its ability to adopt the tetrahedral coordination, are favourable to such a substitution. In contrast to the case of zinc substitution, the maximum amount of lithium that was substituted was limited by the experimental conditions of synthesis. Indeed, working in sealed tubes, and using the precursors $Y_2O_3$, $BaFe_2O_4$, $LiFeO_2$, $Fe_2O_3$ and Fe in order to avoid any reaction with the support and any $Li_2O$ volatization, only the compositions $YBaFe_{4-x}Li_xO_7$ with $x \leq 0.75$ could be prepared, keeping the oxygen and lithium stoichiometry intact.

The first important point deals with the fact that lithium substitution, like zinc, stabilizes the hexagonal symmetry at the cost of the cubic phase. Quite remarkably, a smaller lithium content, $x = 0.30$ only, is sufficient to stabilize the hexagonal form, instead of $x = 0.40$ for Zn. In any case, the cell parameters of the Li substituted yttrium phase vary only slightly with composition from $a = 6.3074$ (1) Å, $c = 10.3586$ (2) Å for $x = 0.30$ to $a = 6.2932$ (1) Å, $c = 10.3104$ (1) Å for $x = 0.75$.

The magnetic study of the compounds $YBaFe_{4-x}Li_xO_7$ clearly shows that for the maximum substitution rate i.e. $x = 0.75$, the complete spin glass behaviour cannot be reached. Thus, for the sake of comparison with other substituted phases, we discuss, in this section, the results obtained in the low doping regime.

### 3.2.1. *$YBaFe_{3.7}Li_{0.3}O_7$ and $CaBaFe_{3.7}Li_{0.3}O_7$*

$Li^+$ substitution in hexagonal $CaBaFe_4O_7$ has been studied by us before [16]. For the purpose of comparison with $Li^+$ in $YBaFe_4O_7$, we choose the samples with $x = 0.3$ from the two series. This is because, as we have stated before, $x = 0.3$ is the minimum amount of $Li^+$ required to stabilize monophasic $YBaFe_{4-x}Li_xO_7$ with the hexagonal symmetry.

The d. c. magnetization results of $YBaFe_{3.7}Li_{0.3}O_7$ have been shown in **Fig. 6**. In accordance with its hexagonal symmetry, $YBaFe_{3.7}Li_{0.3}O_7$ is ferrimagnetic. However, considering the fact that the average Fe valency in YBaFe$_{3.7}$Li$_{0.3}$O$_7$ (Fe$_{val}$ = 2.35) is much less than that in CaBaFe$_{3.7}$Li$_{0.3}$O$_7$ (Fe$_{val}$ = 2.62), the ferrimagnetic interaction in YBaFe$_{3.7}$Li$_{0.3}$O$_7$ should be weaker than that in CaBaFe$_{3.7}$Li$_{0.3}$O$_7$. This is indeed the case as can be seen from **Table 1**, where we have compared the values of T$_C$, M$_{FC(T=5K)}$ and H$_{C(T=5K)}$ for the two samples. The values for CaBaFe$_{3.7}$Li$_{0.3}$O$_7$ in **Table 1** have been quoted from reference 16. CaBaFe$_{3.7}$Li$_{0.3}$O$_7$ has higher values of T$_C$, M$_{FC(T=5K)}$ and H$_{C(T=5K)}$ than those of YBaFe$_{3.7}$Li$_{0.3}$O$_7$, showing that the ferrimagnetic interaction is stronger and the magnetic frustration is weaker in CaBaFe$_{3.7}$Li$_{0.3}$O$_7$.

### 3.2.2. *CaBaFe$_{3.8}$Li$_{0.2}$O$_7$ and CaBaFe$_{3.5}$Zn$_{0.5}$O$_7$*

In all the above cases, we have compared samples with different average Fe valencies, and shown that in the regime of low doping, the Fe valency controls the magnetic properties of the oxide, while in the regime of high doping, the degree of cationic disorder is the deciding factor. We have seen that in the low doping regime, samples with the same degree of disorder, but with different Fe valency behave differently vis – à – vis their magnetic properties. In our final section, we investigate what happens in the reverse case i.e. for samples with different degree of disorder (but well within the low doping regime), but having the same average Fe valency. For this purpose, we compare the two samples CaBaFe$_{3.8}$Li$_{0.2}$O$_7$ and CaBaFe$_{3.5}$Zn$_{0.5}$O$_7$ which fall in the low doping regime, and have almost the same average Fe valency (Fe$_{val}$ = 2.57 and 2.58 for CaBaFe$_{3.8}$Li$_{0.2}$O$_7$ and CaBaFe$_{3.5}$Zn$_{0.5}$O$_7$ respectively).

The d. c. magnetization results of CaBaFe$_{3.5}$Zn$_{0.5}$O$_7$ have been shown in **Fig. 7**. In accordance with its hexagonal symmetry, CaBaFe$_{3.5}$Zn$_{0.5}$O$_7$ shows a sharp ferrimagnetic transition, and a large square hysteresis loop. However, what is more striking is the almost exact one – to – one correspondence of the magnetic parameters obtained for the two samples CaBaFe$_{3.8}$Li$_{0.2}$O$_7$ and CaBaFe$_{3.5}$Zn$_{0.5}$O$_7$. These are listed in **Table 2**. The values for CaBaFe$_{3.8}$Li$_{0.2}$O$_7$ have been quoted from reference 16. This remarkable correspondence in the magnetic properties of the two samples with the same average Fe valency proves unambiguously that in the low doping regime the Fe valency is indeed the main factor controlling the magnetic state of the oxide.

## 4. Conclusion

This study shows the great impact of the substitution of diamagnetic cations such as zinc or lithium for iron upon the competition between ferrimagnetism and magnetic frustration in "114" ferrites. The first effect is structural – it is seen that the substitution of these cations for iron in the cubic phase, $YBaFe_4O_7$, leads to a hexagonal symmetry, and consequently destroys the 3 D geometric frustration at the benefit of a competition between a 2 D geometric frustration and 1 D magnetic ordering, thereby inducing ferrimagnetism. The second effect, observed in the hexagonal phases such as $CaBaFe_4O_7$, modifies the competition between the 2 D frustration and the 1 D magnetic ordering in two different ways, depending on the substituent concentration. For low doping values, the modification of the average iron valency that is induced by this substitution dominates the magnetism of these compounds leading to a decrease of ferrimagnetism at the benefit of 2 D magnetic frustration, whereas in the high doping regime, the disordering of the cations dominates, inducing a complete magnetic frustration, irrespective of the iron valency. The nature of the ferrimagnetism of these ferrites, till date, has not been completely elucidated, and in particular, it is still not known whether the iron spins lie in plane or out of the triangular planes, so that a vast field is still open for the investigation and understanding of this new type of magnetic frustration.

## 5. Acknowledgements

We acknowledge the CNRS and the Conseil Regional of Basse Normandie for financial support in the frame of Emergence Program and N°10P01391. V. P. acknowledges support by the ANR-09-JCJC-0017-01 (Ref: JC09_442369).

**Figure Captions**

**Figure 1**: Structure of (a) cubic LnBaFe$_4$O$_7$ and (b) hexagonal CaBaFe$_4$O$_7$, built up of two sorts of layers of FeO$_4$ tetrahedra called triangular (T) and kagomé (K).

**Figure 2**: Schematic representation of the iron sublattice in (a) cubic LnBaFe$_4$O$_7$ and (b) hexagonal CaBaFe$_4$O$_7$.

**Figure 3**: M$_{ZFC}$(T) and M$_{FC}$(T) curves for YBaFe$_{3.5}$Zn$_{0.5}$O$_7$ and CaBaFe$_{3.5}$Zn$_{0.5}$O$_7$ measured at H = 0.3 T. The inset shows the magnetization as a function of magnetic field at T = 5 K for the two samples.

**Figure 4**: M$_{ZFC}$(T) and M$_{FC}$(T) curves for YBaFe$_{2.5}$Zn$_{1.5}$O$_7$ and CaBaFe$_{2.5}$Zn$_{1.5}$O$_7$ measured at H = 0.3 T. The inset shows the magnetization as a function of magnetic field at T = 5 K for the two samples.

**Figure 5**: Real (in-phase) component of a.c. susceptibilities for (a) YBaFe$_{2.5}$Zn$_{1.5}$O$_7$ and (b) CaBaFe$_{2.5}$Zn$_{1.5}$O$_7$ as a function of temperature measured using a frequency range 10 Hz – 10 kHz.

**Figure 6**: M$_{ZFC}$(T) and M$_{FC}$(T) curves for YBaFe$_{3.7}$Li$_{0.3}$O$_7$ measured at H = 0.3 T. The insets (a) show the magnetization as a function of magnetic field at T = 5 K and (b) dM/dT as a function of T for estimation of T$_C$.

**Figure 7**: M$_{ZFC}$(T) and M$_{FC}$(T) curves for CaBaFe$_{3.5}$Zn$_{0.5}$O$_7$ measured at H = 0.3 T. The insets (a) shows the plot of d. c. magnetic susceptibility as a function of temperature along with the Curie – Weiss fit, (b) the magnetization as a function of magnetic field at T = 5 K and (c) dM/dT as a function of T for estimation of T$_C$.

**Table Captions**

**Table 1**: T$_C$, M$_{FC(T=5K)}$ and coercive field (H$_C$) for YBaFe$_{3.7}$Li$_{0.3}$O$_7$ and CaBaFe$_{3.7}$Li$_{0.3}$O$_7$.

**Table 2**: T$_C$, Curie-Weiss temperature ($\theta_{CW}$), effective paramagnetic moment ($\mu_{eff}$) and coercive field (H$_C$) for CaBaFe$_{3.8}$Li$_{0.2}$O$_7$ and CaBaFe$_{3.5}$Zn$_{0.5}$O$_7$.

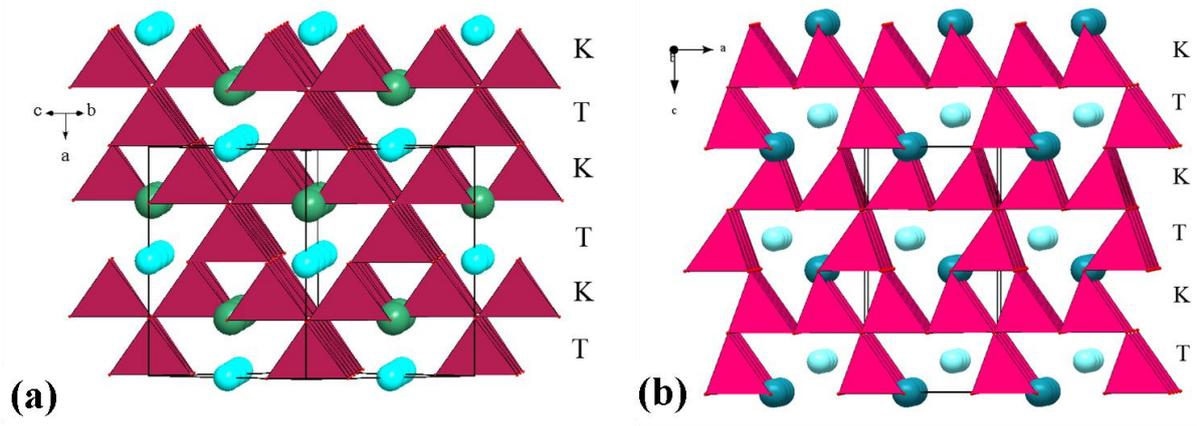

Fig. 1. Structure of (a) cubic LnBaFe$_4$O$_7$ and (b) hexagonal CaBaFe$_4$O$_7$, built up of two sorts of layers of FeO$_4$ tetrahedra called triangular (T) and kagomé (K).

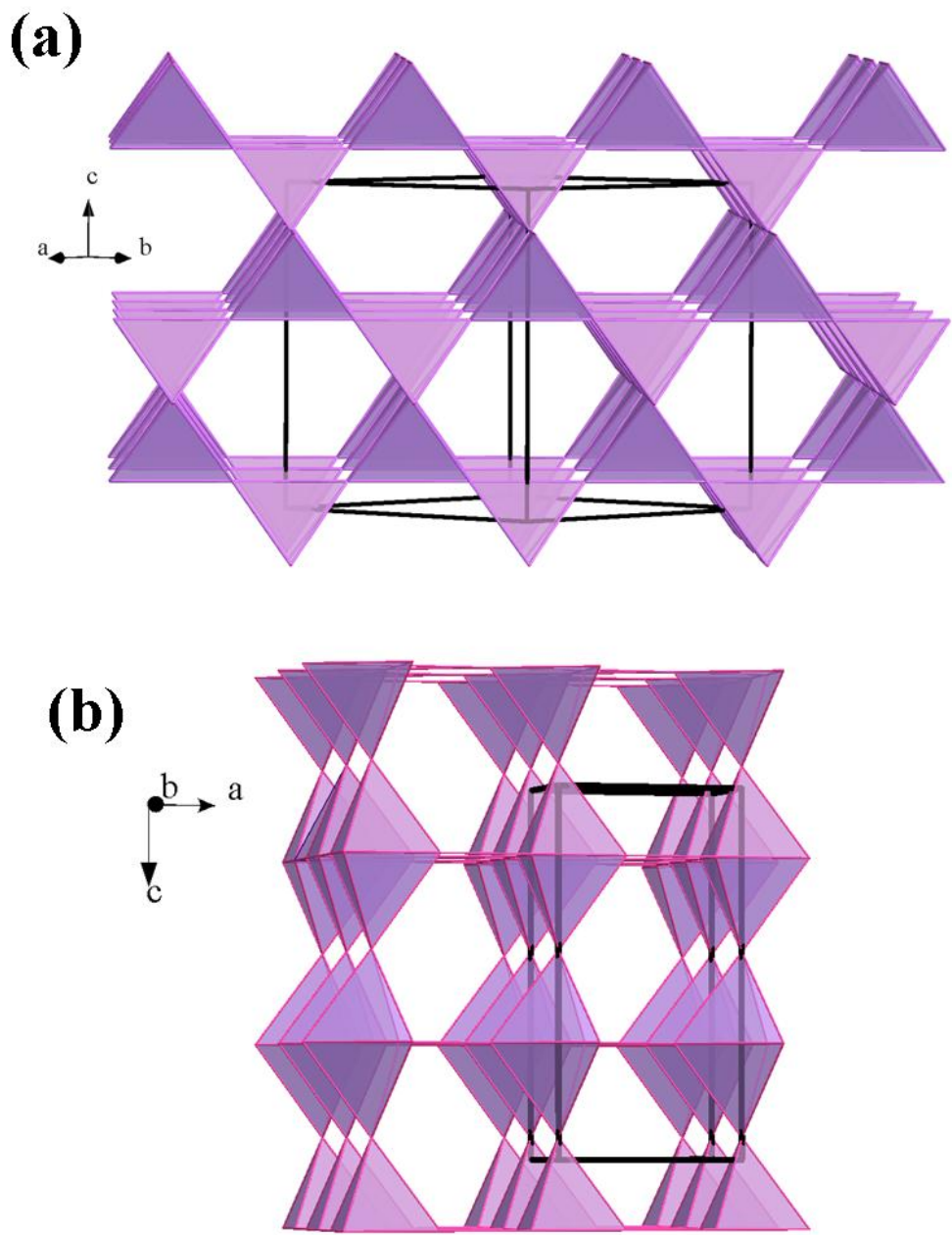

Fig. 2. Schematic representation of the iron sublattice in (a) cubic LnBaFe$_4$O$_7$ and (b) hexagonal CaBaFe$_4$O$_7$.

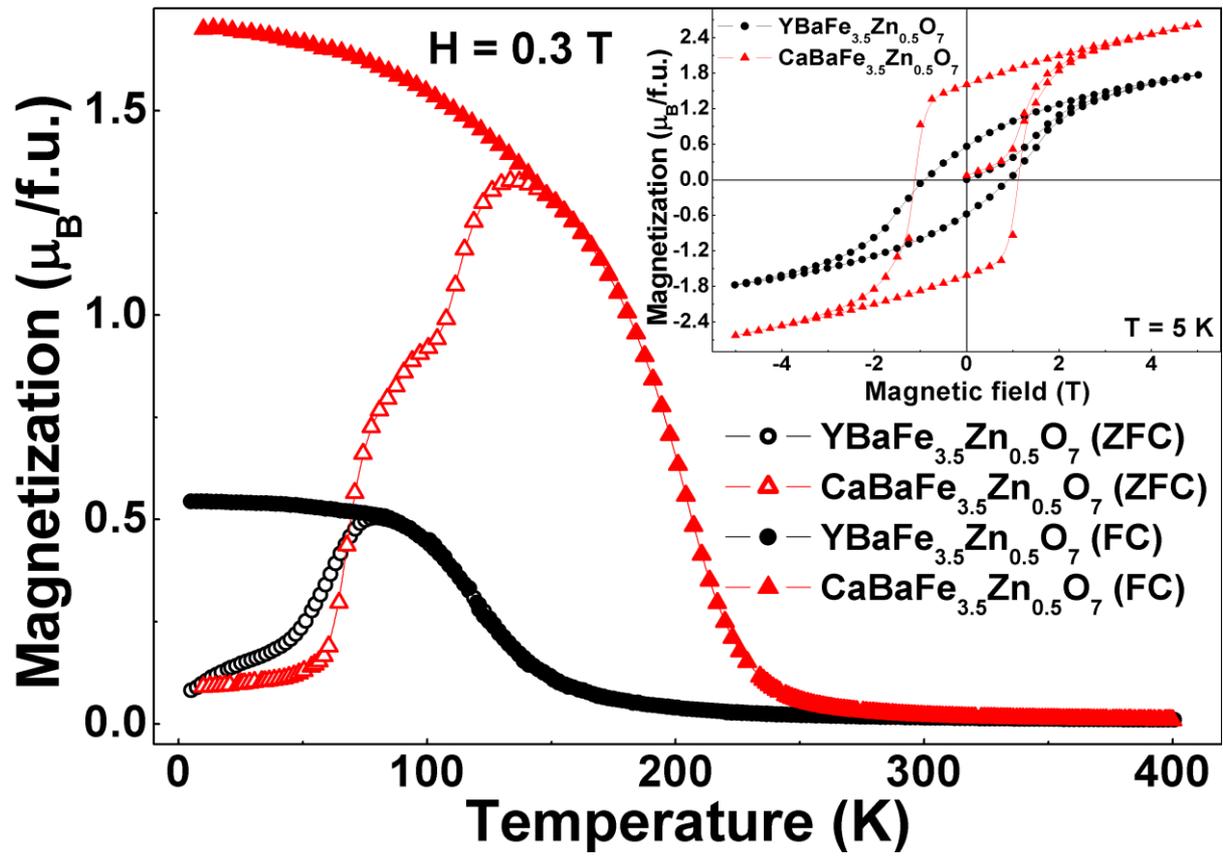

Fig. 3. $M_{ZFC}(T)$ and $M_{FC}(T)$ curves for $YBaFe_{3.5}Zn_{0.5}O_7$ and $CaBaFe_{3.5}Zn_{0.5}O_7$ measured at H = 0.3 T. The inset shows the magnetization as a function of magnetic field at T = 5 K for the two samples.

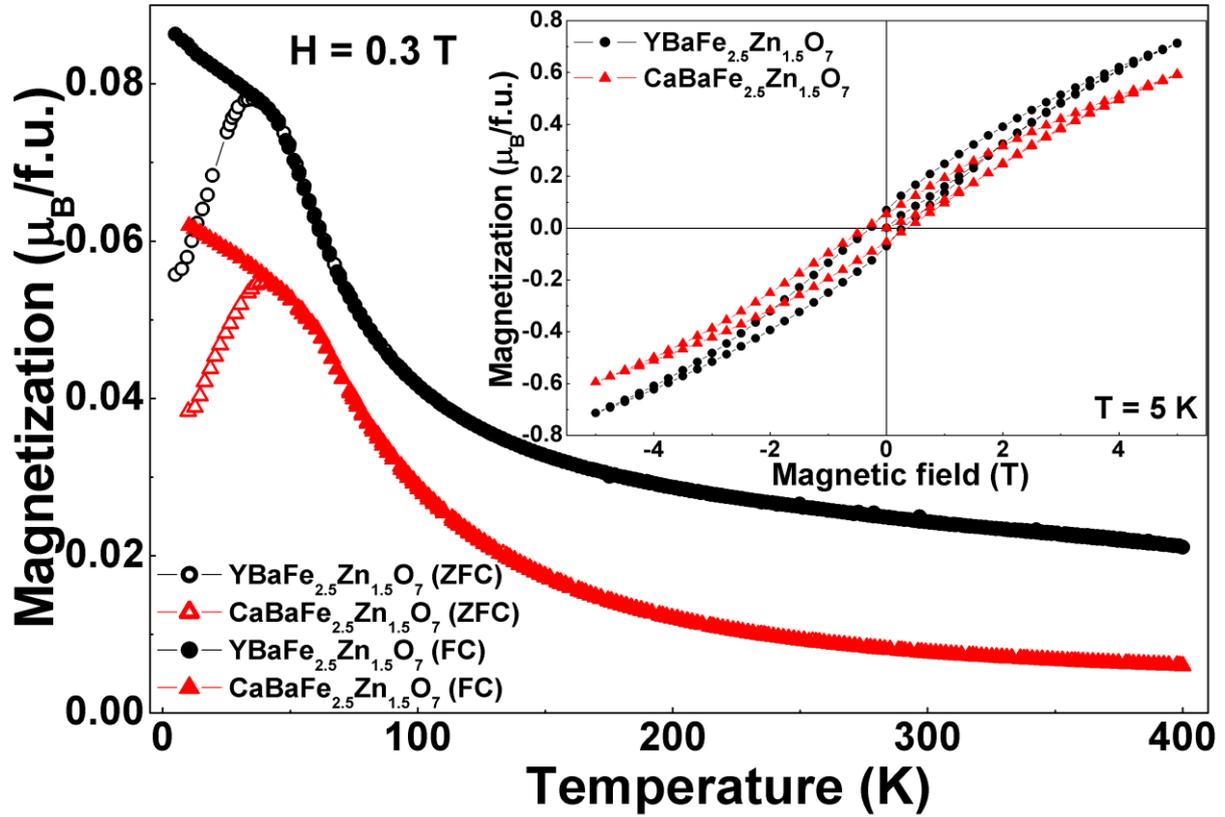

Fig. 4. $M_{ZFC}(T)$ and $M_{FC}(T)$ curves for $YBaFe_{2.5}Zn_{1.5}O_7$ and $CaBaFe_{2.5}Zn_{1.5}O_7$ measured at H = 0.3 T. The inset shows the magnetization as a function of magnetic field at T = 5 K for the two samples.

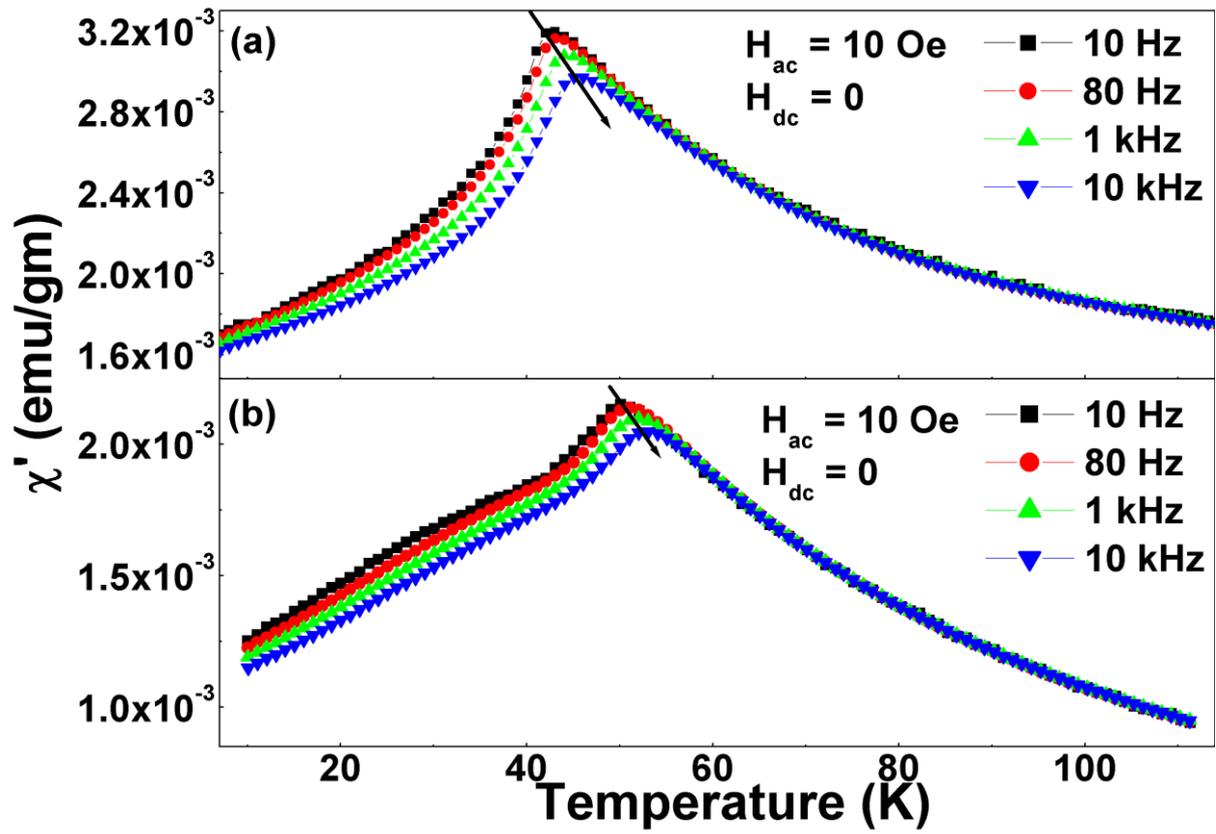

Fig. 5. Real (in-phase) component of a.c. susceptibilities for (a) $YBaFe_{2.5}Zn_{1.5}O_7$ and (b) $CaBaFe_{2.5}Zn_{1.5}O_7$ as a function of temperature measured using a frequency range 10 Hz – 10 kHz.

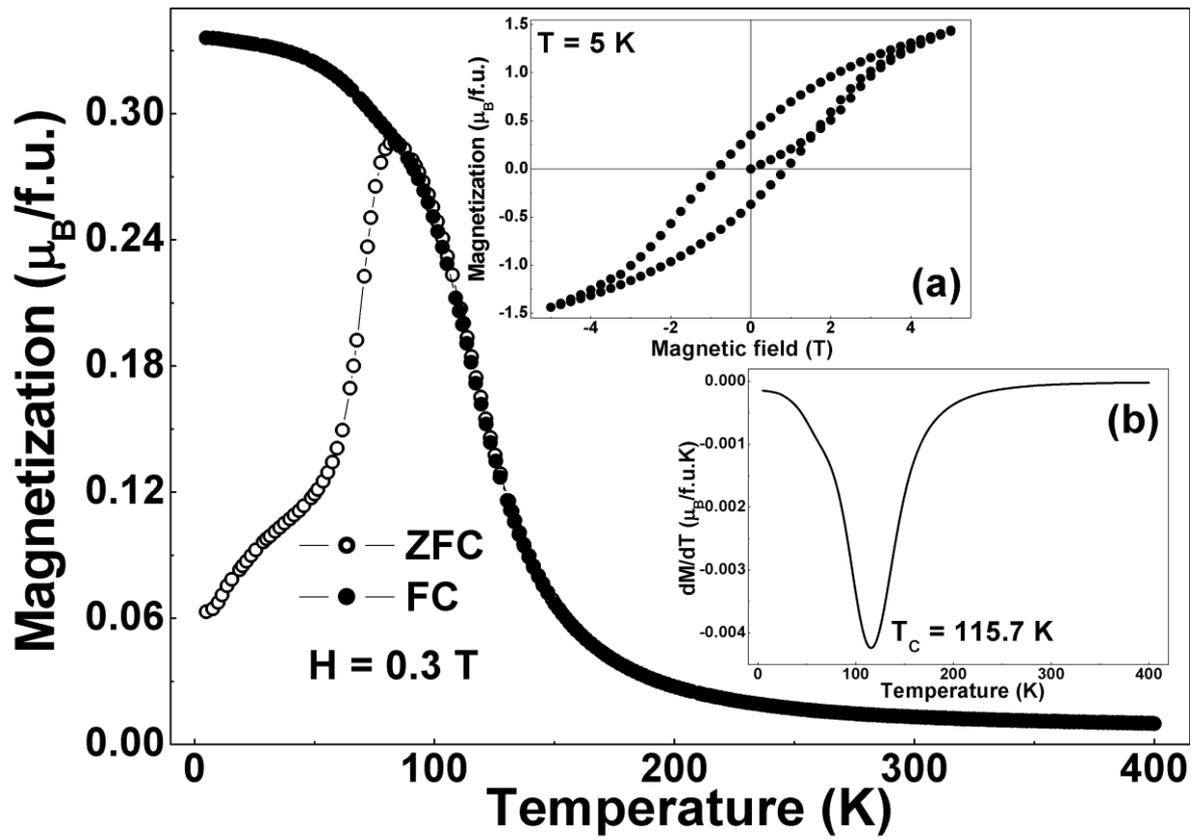

Fig. 6. $M_{ZFC}(T)$ and $M_{FC}(T)$ curves for YBaFe$_{3.7}$Li$_{0.3}$O$_7$ measured at H = 0.3 T. The insets (a) show the magnetization as a function of magnetic field at T = 5 K and (b) dM/dT as a function of T for estimation of T$_C$.

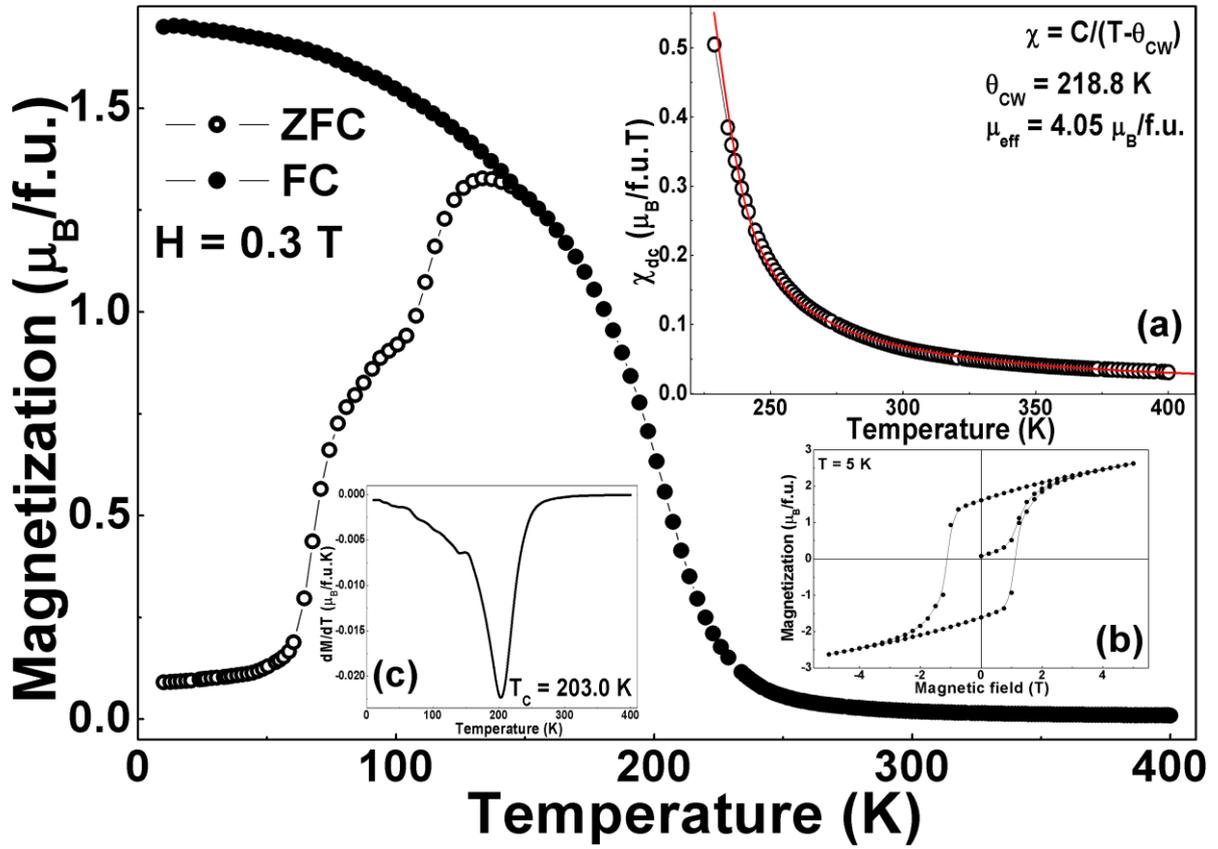

Fig. 7. $M_{ZFC}(T)$ and $M_{FC}(T)$ curves for $CaBaFe_{3.5}Zn_{0.5}O_7$ measured at H = 0.3 T. The insets (a) shows the plot of d. c. magnetic susceptibility as a function of temperature along with the Curie – Weiss fit, (b) the magnetization as a function of magnetic field at T = 5 K and (c) dM/dT as a function of T for estimation of $T_C$.

**Table 1**. $T_C$, $M_{FC(T=5K)}$ and coercive field ($H_C$) for $YBaFe_{3.7}Li_{0.3}O_7$ and $CaBaFe_{3.7}Li_{0.3}O_7$.

| Sample | $T_C$ (K) | $M_{FC(T=5K)}$ ($\mu_B$/f.u.) | $H_C$ (at T = 5 K) (T) |
|---|---|---|---|
| $YBaFe_{3.7}Li_{0.3}O_7$ | 115.7 | 0.34 | 0.87 |
| $CaBaFe_{3.7}Li_{0.3}O_7$ | 150.7 | 0.50 | 0.98 |

**Table 2**. $T_C$, Curie-Weiss temperature ($\theta_{CW}$), effective paramagnetic moment ($\mu_{eff}$) and coercive field ($H_C$) for $CaBaFe_{3.8}Li_{0.2}O_7$ and $CaBaFe_{3.5}Zn_{0.5}O_7$.

| Sample | $T_C$ (K) | $\theta_{CW}$ (K) | $\mu_{eff}$ ($\mu_B$/f.u.) | $H_C$ (at T = 5 K) (T) |
|---|---|---|---|---|
| $CaBaFe_{3.8}Li_{0.2}O_7$ | 191.6 | 211.9 | 4.37 | 1.17 |
| $CaBaFe_{3.5}Zn_{0.5}O_7$ | 203.0 | 218.8 | 4.05 | 1.12 |